# An Experimentally Driven Automated Machine Learned Inter-Atomic Potential for a Refractory Oxide

Ganesh Sivaraman*

Leadership Computing Facility, Argonne National Laboratory, Argonne, IL 60439, USA

Leighanne Gallington

X-ray Science Division, Argonne National Laboratory, Argonne, IL 60439, USA

Anand Narayanan Krishnamoorthy

Helmholtz-Institute Munster: Ionics in Energy Storage (IEK-12),

Forschungszentrum Julich GmbH, Corrensstrasse 46, 48149 Munster, Germany

Marius Stan

Applied Materials Division, Argonne National Laboratory, Argonne, IL 60439, USA

Gabor Csanyi

Department of Engineering, University of Cambridge,

Trumpington Street, Cambridge, CB2 1PZ, United Kingdom

Alvaro Vazquez-Mayagoitia

Computational Science Division, Argonne National Laboratory, Argonne, IL 60439, USA

Chris J. Benmore[†]

X-ray Science Division, Argonne National Laboratory, Argonne, IL 60439, USA

* Also at Data Science and Learning Division, Argonne National Laboratory, Argonne, IL 60439, USA

[†] benmore@anl.gov

**Abstract.**

Understanding the structure and properties of refractory oxides are critical for high temperature applications. In this work, a combined experimental and simulation approach uses an automated closed loop via an active-learner, which is initialized by X-ray and neutron diffraction measurements, and sequentially improves a machine-learning model until the experimentally predetermined phase space is covered. A multi-phase potential is generated for a canonical example of the archetypal refractory oxide, $HfO_2$, by drawing a minimum number of training configurations from room temperature to the liquid state at ~2900°C. The method significantly reduces model development time and human effort.

Refractory oxides are essential components in the development of high temperature ceramic materials [1], thermal barrier coatings [2] and nuclear applications [3,4]. Their high melting temperatures, $T_m$>1500°C, make refractories suitable for applications in harsh environments, in addition to their insulating properties and ability to prevent oxidation. It is therefore important to identify phase transformations and structural rearrangements close to the melting point. Diffraction plays an important role in the computing of phase diagrams and thermochemistry using the CALPHAD method, which has been the foundation for providing a consistent picture of the stable structures and thermodynamic properties of materials through the calculation of the Gibbs free energy. X-ray powder diffraction in particular is a workhorse for materials characterization, providing data on crystallographic phases, thermal expansion and volume changes associated with phase transitions in different atmospheres. Neutron powder diffraction also provides valuable structural information, especially on lighter elements such as oxygen, but generally requires larger samples and longer count times. However there are few suitable containers for X-ray and neutron diffraction experiments at temperatures >2000°C. In the last decade, advances in aerodynamic levitation and laser heating techniques combined with high-energy X-ray and neutron diffraction have pushed crystallographic measurements above 1500°C [5,6] providing accurate structural data over a wide range of phase space.

On the computational modeling front, *Ab initio* Molecular Dynamics Simulations (AIMD) provide atomic scale resolution with quantum mechanical accuracy, but are restricted to short simulation times, and small system sizes. Empirical inter-atomic potentials based on fixed analytical functional forms are derived from physical or chemical intuitions and parametrized to experimental properties, but lack the sophistication to capture the many-body interactions required to arrive at *ab intio* accuracies. In recent years, advances in combining quantum-mechanical calculation calculations with machine learning has resulted in a new class of inter-atomic potentials that learns the potential energy surface landscape directly from reference *ab initio* datasets [7–11]. Machine learning inter-atomic potentials (ML-IP) can maintain near *ab initio* accuracy while affording atomic resolution at larger system sizes (through linear scaling) and time scales comparable to classical inter-atomic potentials [12,13,56]. In particular ML-IP based on the Gaussian Approximation Potential (GAP) [14] have been successfully applied to model liquids [15,16], crystals [17], defects [16], amorphous [18], multi-component materials [19] and molecules [20]. Training ML-IP requires efficiently drawing configurations from a wide chemical space of interest and finding the best hyper-parameters. Active learning is a sub-domain of machine learning where an unsupervised machine learning arrives at an optimal supervised machine learning model (i.e. ML-IP) with a minimum number of training configurations [21]. Smith *et al.* proposed a "query by committee" strategy, which is an active learning strategy that exploits disagreement in ensemble of ML-IP model by sampling regions of chemical space where the ML-IP fails to predict the potential energy accurately [22]. Podryabinkin *et al.* used an active learning strategy based on a "D-optimality" criterion for selecting atomic configurations [23]. Zhang *et al.* employed a Deep Potential Generator to efficiently sample configuration space, and generate an accurate reference dataset from the configuration with low prediction accuracy, and perform iterative training [24]. Active learning strategies based on Bayesian inference have also been reported [19,25,26]. We recently reported an active learner that relied on exploiting the cluster structure embedded in a given unlabeled atomic configurations so as to arrive at a minimum number of training configurations [15,27]. Here we propose to bring together the advances in experiments at extreme conditions and theoretical modeling through a closed loop active learning scheme as shown in Figure 1. Our scheme consists of three components: (1) Experimental measurements are performed up to the melting temperature on a refractory oxide sample. Model structures are fitted to the neutron and X-ray diffraction measurements of each of the phases at different reference temperatures. *In-situ* high energy X-ray diffraction is used to obtain unit cell volume as a function of temperature. (2) An active learning scheme initialized by the model structures drives the phase space exploration over the experimental measurement region. (3) A ML-IP is generated that can be iteratively improved by the active learning scheme. To illustrate this approach, we consider an archetypal refractory oxide, Hafnium dioxide, $HfO_2$ (which is isostructural with the most studied ceramic $ZrO_2$). Upon heating $HfO_2$ undergoes transformations from monoclinic (m-$HfO_2$) to tetragonal (t-$HfO_2$) to cubic (c-$HfO_2$) phases before melting at ~2800°C [28,29].

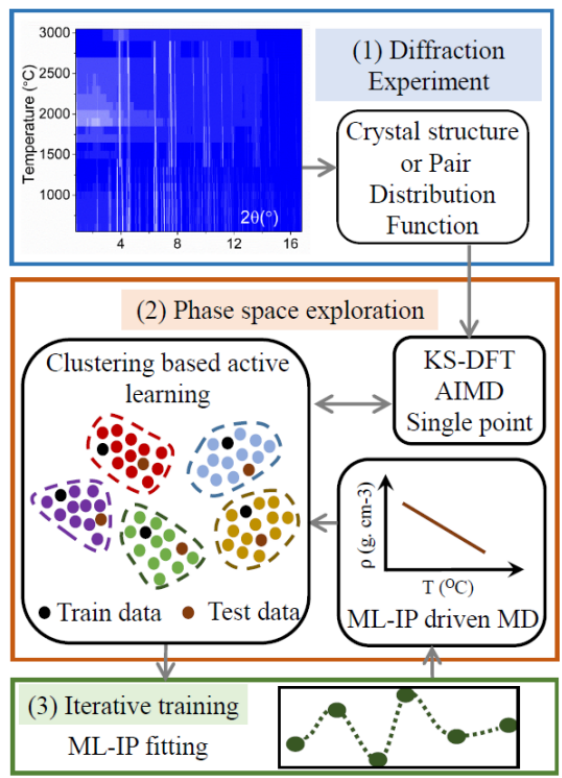

FIG. 1. The experiment driven workflow. (1) Experimental high energy X-ray and neutron diffraction patterns are measured over a wide temperature range using a uni-axial laser heating system on an aerodynamically levitated $HfO_2$ sample. (2) Cluster based active learning enables exploration over a wide range of phase space (3) Iterative training and fitting methods provides feedback into (2).

The experiment driven workflow is shown in Figure 1. X-ray diffraction data were collected at beamline 6-ID-D at the Advanced Photon Source, Argonne National Laboratory on an amorphous silicon area detector (PE-XRD1621) using 60.07 keV (λ=0.2064Å) X-rays. High purity samples (Aldrich, 99.995% trace metal purity) of ~2mm diameter were levitated and heated up to ~3000°C in reducing (argon) and oxidizing (oxygen) atmospheres [29]. Calibration of the detector distance, beam center, detector tilt and rotation were performed using the *Fit2D* software package based on the measurement of a $CeO_2$ NIST standard [30]. Reduction of the 2-D images to 1-D diffraction patterns yielded the X-ray intensities, $I_{XRAY}(Q)$. Lattice parameters were obtained via LeBail whole pattern fitting of the previously reported monoclinic ($P2_1/c$), tetragonal ($P4_2/nmc$), and cubic ($Fm\text{-}3m$) crystal structure models to the diffraction data [31–33]. The volumes obtained were normalized to the number of $HfO_2$ formula units per unit cell to aid in the comparison of the cubic and monoclinic unit cell (Z=4) volumes to that of the tetragonal phase (Z=2). The phase transitions from monoclinic-tetragonal-cubic-liquid in an Argon atmosphere with increasing temperature are shown in Figure 2. A deviation of ~0.2% to lower V is observed in an oxygen atmosphere for the monoclinic phase for temperatures >600°C and the phase transition to tetragonal occurs at 1400-1500°C depending on redox environment [34]. However, the cubic and tetragonal phase volumes are essentially the same in both Ar and $O_2$ [34]. Figure 3(a) and 3(b) shows the X-ray data for the high temperature crystalline phases and the liquid and amorphous forms. The latter $S_{XRAY}(Q)$ experimental data have previously been reported in [29] but are shown here to show the extent to which this multi-phase potential has been trained.

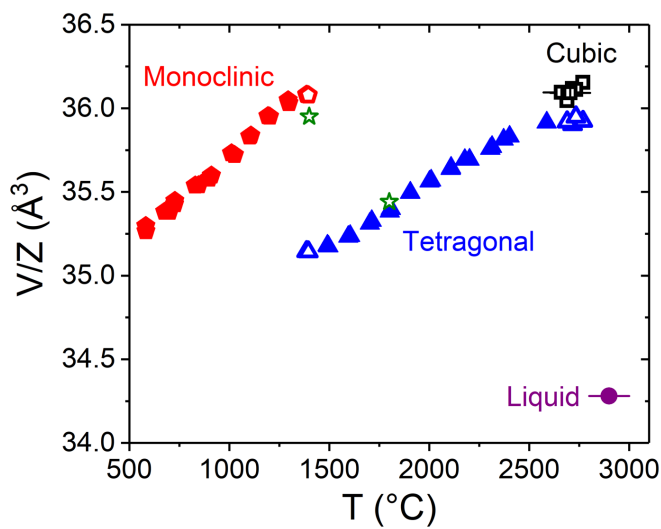

FIG. 2. Unit cell volume of the monoclinic (pentagon), tetragonal (triangles) and cubic (squares) forms of hafnia measured in an Argon atmosphere. Open symbols represent mixed phases and solid symbols are single phase. The cubic form was only observed as a mixed phase with the tetragonal polymorph. The average cell volume estimates from GAP MD simulation for m-$HfO_2$ at 1400°C and 1800°C with in an isothermal-isobaric ensemble are shown as green stars.

Complementary neutron diffraction measurements were performed on the NOMAD beamline at the Spallation Neutron Source (Oak Ridge National Laboratory). Data were acquired for each of the crystalline phases of $HfO_2$ i.e. monoclinic at T~1000°C, tetragonal at ~1850°C and cubic at ~2900°C in Argon and in a 80%Ar:20%$O_2$ mixture using a laser-heated aerodynamic levitator [35]. The time-of-flight neutron data were reduced using in-house software [36] to extract the pair distribution functions, $G_{NEUTRON}(r)$. Neutron pair distribution functions for hafnia in the monoclinic, tetragonal and cubic+tetragonal forms. Neutron levitation experiments are considerably more difficult that X-rays due to the lower signal/background ratio and long count times required. However, neutrons are more sensitive to oxygen correlations than X-rays which are important for understanding defects and diffusion i.e. at Q=0 Å$^{-1}$ the O-O neutron partial weighting factor is 36% compared to 3% for X-rays, and the Hf-O partial is 48% (neutrons) compared to 30% (X-rays).

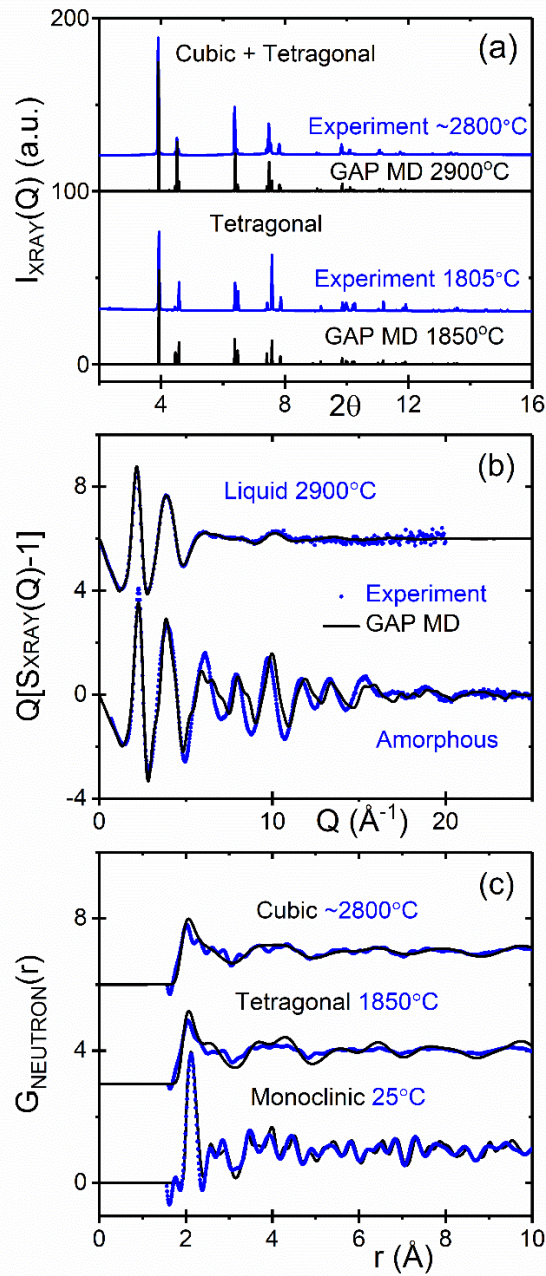

FIG. 3. Diffraction versus simulation data for (a) X-ray diffraction patterns compared to the GAP MD computed X-ray intensity ($\lambda$=0.12359 Å) for the two high temperature phases. (b) Experimental and simulated X-ray structure factors for amorphous and liquid $HfO_2$ (c) Experimental and simulated neutron pair distribution functions for the pure phases of $HfO_2$

The second section in the workflow illustrated in figure 1 is the experimentally driven phase space exploration: Here we implement in the closed-loop, active learning, phase space exploration in two steps. In the initialization step, active learning starts from the model structures, and generates the ML-IP model from the ensemble of AIMD structures at the neutron diffraction reference temperatures. This initial ML-IP model corresponds to a poor approximation of the potential energy surface and is used to perform ensemble of isothermal-isobaric molecular dynamics corresponding to regions unexplored in the experiment. The active learning sub-samples data from those MD simulation trajectories to perform *ab initio* single point calculation and iteratively retrain the ML-IP model. The active learning exits the closed loop once the required phase space coverage is achieved. The active learning phase space exploration region corresponds to heating m-$HfO_2$ from 25 to 1500°C, cooling t-$HfO_2$ from 1850 to 1400°C, heating t-$HfO_2$ from 2400 to 1850°C and cooling c-$HfO_2$ from 2900 to 2300°C.

The active learning process is built on a recently proposed scheme based on an unsupervised clustering method coupled to a Bayesian Optimization (BO) [15]. The unsupervised clustering method uses the HDBSCAN algorithm to partition the input trajectory and sequentially samples sparse configurations for training the ML-IP [37,38]. The Bayesian optimization performs on-the-fly hyper-parameter optimizations, to find the optimal ML-IP model by training on the sampled configurations and validating an independently sampled test dataset [39]. The advantage of this approach is that BO also provides the optimal hyper-parameters on-the-fly. Previously [39], the active learning method has been applied to a very large AIMD `melt-quench` dataset of 33,000 configurations. Here, we show that the scheme is able to arrive at a near *ab initio* accurate training dataset with only 0.8% i.e. 260 samples from this large dataset. For the ML-IP, we use the Gaussian Approximation Potential (GAP) model along with the many-body Smooth Overlap of Atomic Positions (SOAP) descriptor [14,40]. The details of the GAP model and the descriptor are further discussed in supplementary material section B [34]. The active learning scheme is further discussed in the supplementary material section C [34] and the code implementation with examples usage with GAP model are available elsewhere [15].

The *ab initio* calculations were performed using Density Functional Theory (DFT) as implemented in *VASP* package [41,42]. The Perdew-Burke-Ernzerhof generalized gradient approximation and projector augmented plane wave methods were employed [43,44], with a 520eV planewave cutoff and 2x2x2 K-grid. A 1 fs time step and Nosé–Hoover thermostat were used for the AIMD [35,36]. For the iterative training of the GAP model a system size of 96-atom was employed, except for t-HfO2 where a 108-atom system size was used. An ensemble of AIMD simulations were performed for 12 ps starting from the pure phases based on the model structures at the neutron diffraction reference temperatures. The active learning was initialized with the last 6000 snapshots from the each of the AIMD trajectories. Using the initial active learned GAP model, isothermal-isobaric ensemble sampling was performed with the *LAMMPS* simulation package compiled with the QUIP pair style support [45–47]. The single point DFT calculations employ the same DFT parameters as discussed above. The training dataset generation and simulation set up for amorphous and liquid $HfO_2$ have been discussed in detail elsewhere [15]. Here the simulation for both the liquid and amorphous forms have been recalculated with the multi-phase potential to show the entire phase space.

The training of the ML-IP based on the GAP model was iteratively mapped by the active learning, until a uniform coverage is achieved across experimental phase space. The active-learned multi-phase potential provides a model that spans the entire phase space regions from the liquid to amorphous and crystalline states of $HfO_2$ with a meagre 2053 configurations, the details of which are summarized in supplementary material Table. C1 [34]. The parameters used for training the multi-phase potential are summarized in the supplementary material Table. C2 [34]. A non-parametric two-body term was added to the SOAP descriptor to prevent unphysical clustering of atoms at high temperatures [48]. The multi-phase potential was validated on a randomly drawn DFT configuration (i.e. outside of training dataset) from the entire phase space region and gave a mean absolute error in energy of 2.4 meV/atom.

For the GAP MD based production simulation, a 6144-atom simulation cell was used for both m-$HfO_2$ and c-$HfO_2$. For t-$HfO_2$, 6912-atom system was used. The trajectories were sampled for 1.1 ns at the reference neutron diffraction temperatures. The first 100 ps were omitted and the subsequent 1 ns trajectory was used for the analysis. The structural arrangements for all the phases of $HfO_2$ are shown in Figure 3. Figure 3(a) shows good agreement between the Hf-dominated experimental X-ray intensities and GAP MD simulations for the two high temperature crystalline phases [32,33,49]. Similarly, figure 3(c) compares the O-sensitive neutron diffraction patterns for the monoclinic, tetragonal and cubic forms of $HfO_2$. The structure factors of the simulated liquid and amorphous form are shown in figure 3(b). Long-range ordering was found to be diminished considerably with increasing temperature in both the tetragonal and cubic forms, and this increased disorder at high temperatures is captured by the machine learned GAP model. The effect of both disorder and density is also seen in the S(Q)'s for the amorphous and liquid phases. Here strong oscillations in the lower density amorphous signal at high- Q (~5-15 $Å^{-1}$) correspond to the edge/corner sharing ratio but these are washed out in the liquid signal. The computation of PDF, structure factor and X-ray intensities are further discussed in supplementary material sections F and H.

In order to assess the quality of the reported multi-phase potential with two well-known parametrizations for $HfO_2$, a comparison of cohesive energy and diffusion coefficients are presented. Since the focus is on experiments, the theoretical validation of multi-phase potential is restricted to this comparison. The cohesive energies of m-$HfO_2$, t-$HfO_2$, and c-$HfO_2$ computed by different methods are shown in Table I. The DFT computed cohesive energies reproduce the correct phase order of the phases. The GAP predicted cohesive energy shows the closest agreement with DFT, followed by Charge-optimized many-body potential (COMB) parametrized for the hafnium/hafnium-oxide system [50]. The well-known (classical MD) parametrization for $HfO_2$ by Broglia *et al*. [51], shows a large deviation with respect to DFT. To further test the quality of the multi-phase potential, 50 random configurations from the m-$HfO_2$ AIMD trajectory were drawn and the forces are computed using our method, compared to COMB and Broglia *et al*. The resulting force validation plot with respect to DFT is shown in supplementary material Figure E1 [34], and indicates that the GAP (0.09 eV/ Å) gives the lowest root mean square error in predicted forces with *ab intio* accuracy, significantly outperforming COMB (4.23 eV/ Å) and Broglia *et al*. (10.85 eV/ Å) inter-atomic potentials.

| Method | m-HfO$_2$ | t-HfO$_2$ | c-HfO$_2$ |
|---|---|---|---|
| DFT | -30.52 | -30.30 | -30.18 |
| GAP | -30.51 | -30.34 | -30.27 |
| COMB [50] | -30.69 | -30.51 | -30.40 |
| Broglia *et al*. [51] | -48.05 | -47.79 | -47.77 |

TABLE 1. Cohesive energy (eV/HfO$_2$) for HfO$_2$ computed with different methods.

The diffusion constants were calculated from our molecular dynamics simulations via the mean square displacements of atoms using:

$$D_\alpha = \lim_{t \to \infty} \frac{1}{6t} \langle [\, r_{i,\alpha}(t) - r_{i,\alpha}(0)]^2 \rangle \quad (1)$$

Where $D_\alpha$ is the diffusion constant for the atomic species $\alpha$, $r_i(t)$ denotes the position of the atomic species at time t. Our simulation results for m-HfO$_2$ and t-HfO$_2$ structures show negligible diffusion at simulation temperatures (25,1850 °C) for both Hf and O. Furthermore, we find the diffusion constants of c-HfO$_2$ to be $D_{Hf}$ = 0.12±0.002x10$^{-6}$ cm$^2$/sec for Hf and $D_O$=1.53±0.005x10$^{-5}$ cm$^2$/sec for O. Similarly, for liquid HfO$_2$, the diffusion constants yield $D_{Hf}$ = 3.3796±0.1×10$^{-5}$cm$^2$/s and $D_O$ = 6.2971±0.1×10$^{-5}$cm$^2$/s respectively. These values are in good agreement with previous simulation results reported by Hong et al. [52]. Furthermore, c-HfO$_2$ shows negligible diffusion for Hf compared to O. However, with increasing temperature we observe a strong diffusion of Hf atoms in liquid HfO$_2$ [15], comparable to that of O atoms.

Generating the solid-liquid phase diagram is a rigorous benchmark for testing the capability of the multi-phase inter-atomic potential and predicting the ultra-high temperature region [17]. To this end, we predict the melting point of HfO$_2$ using the Z-method and modified-Z (M-Z) methods [53–55]. The simulations were performed in a microcanonical (NVE) ensemble with a fixed box size and the complete process was performed along the isochore curve. Beginning with a 4x4x4 c-$HfO_2$ (768 atoms) box of fixed volume of V=35.82 $Å^3$, simulations were performed in an NVE ensemble for different initial temperatures that corresponded to different total energies. The results are shown in figure 4 where one branch of the curve contains the solid line, while the other branch represents the liquid line. The lowest temperature at which the liquid exists is the melting point, and highest temperature for the solid corresponds to the supercritical temperature. For the M-Z method we used a 4x4x40 rectangular parallelepiped simulation box for c-$HfO_2$ with 7680 atoms, which is anisotropic in Z-direction. The isochore curve for the M-Z method is known to be dependent on the size of the simulation cell in the Z-direction and this influence have been reported previously [55]. The minima of the liquid branch for the Z-method corresponds to an underestimated melting point of ~2727°C, whereas the M-Z method gives ~ 2826.85°C, which is within 1% of experiment melting point (~2800°C) [28,29]. Given that the GAP model accurately captures the structure of liquid $HfO_2$ with a unit cell V = 42.827 $Å^3$ (see figure 3(b)), we find the change in the unit cell volume going from cubic to liquid melt is $\Delta V \approx 7 Å^3$.

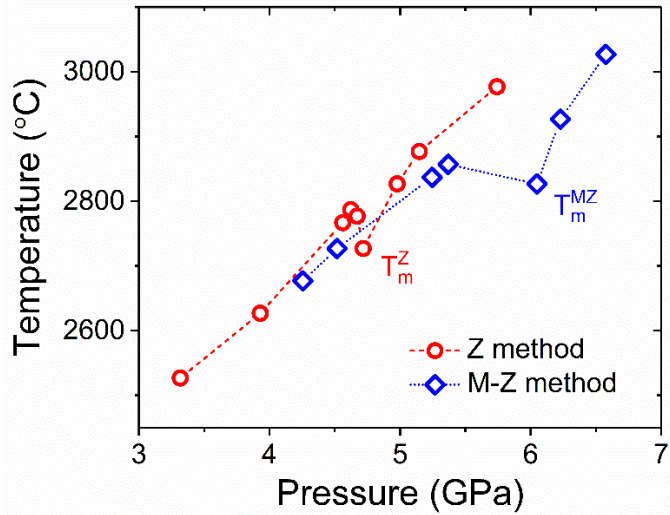

FIG. 4. Isochore curve for V=35.82 $Å^3$.

In conclusion, we show the proof of concept for an automated experimentally driven scheme for generating a multi-phase ML-IP for a canonical refractory oxide namely, $HfO_2$. The approach offers the following distinct advantages: (1) The process removes the ambiguity of sampling phase space required to train the ML-IP's by direct interfacing with experimental measurements. (2) It provides a direct validation of the model with experimental measurement. (3) It also enables experimental model structures to directly enter the training process. (4) Active learning ensures that sparse configurations are required to arrive at an ML-IP within *ab initio* accuracy. The results indicate the multi-phase potential is able to reproduce both the structural and dynamical properties of $HfO_2$ from room temperature to the melt with *ab initio* accuracy. The accuracy of the simulated results are only limited by the choice of the *ab initio* method used for generating the training data and can be systematically improved by choosing more accurate quantum chemistry techniques. Although for this particular application the method involved the Gaussian Approximation Potential framework for generating the model, the proposed scheme could be generalized to other ML-IP methods. Finally, the automation scheme offers a systematic pathway for investigation other refractory oxides and similar classes of materials.

**Acknowledgements.** This material is based upon work supported by Laboratory Directed Research and Development (LDRD) funding from Argonne National Laboratory, provided by the Director, Office of Science, of the U.S. Department of Energy under Contract No. DE-AC02-06CH11357. This research used resources of the Argonne Leadership Computing Facility, which is a DOE Office of Science User Facility supported under Contract DE-AC02-06CH11357. Argonne National Laboratory's work was supported by the U.S. Department of Energy, Office of Science, under contract DE-AC02-06CH11357. We gratefully acknowledge the computing resources provided on Bebop; a high-performance computing cluster operated by the Laboratory Computing Resource Center at Argonne National Laboratory. This research used resources of the Advanced Photon Source, a U.S. Department of Energy (DOE) Office of Science User Facility operated for the DOE Office of Science by Argonne National Laboratory under Contract No. DE-AC02-06CH11357 and the Spallation Neutron Source operated by Oak Ridge National Laboratory. Use of the Center for Nanoscale Materials, an Office of Science user facility, was supported by the U.S. Department of Energy, Office of Science, Office of Basic Energy Sciences, under Contract No. DE-AC02-06CH11357. G.S. would like to thank Dr. Felix Cosmin Mocanu for fruitful discussions on fitting the Gaussian Approximation Potential. A.N.K gratefully acknowledges useful discussions with Prof. Dr. Christian Holm, Dr. Jens Smiatek, Dr. Frank Uhlig, and financial support from the German Funding Agency (Deutsche Forschungsgemeinschaft-DFG) under Germany's Excellence Stratergy - EXC 2075 - 390740016. Rick Weber is thanked for providing the hafnia samples and Joerg Neuefeind is thanked for his help with the neutron experiments.

Supplementary Material for:

# An Experimentally Driven Automated Machine Learned Inter-Atomic Potential for a Refractory Oxide.

Ganesh Sivaraman, Leighanne Gallington, Anand Narayanan Krishnamoorthy, Marius Stan, Gabor Csanyi, Alvaro Vazquez-Mayagoitia, Chris J. Benmore.

## A. Unit cell volume of pure phases.

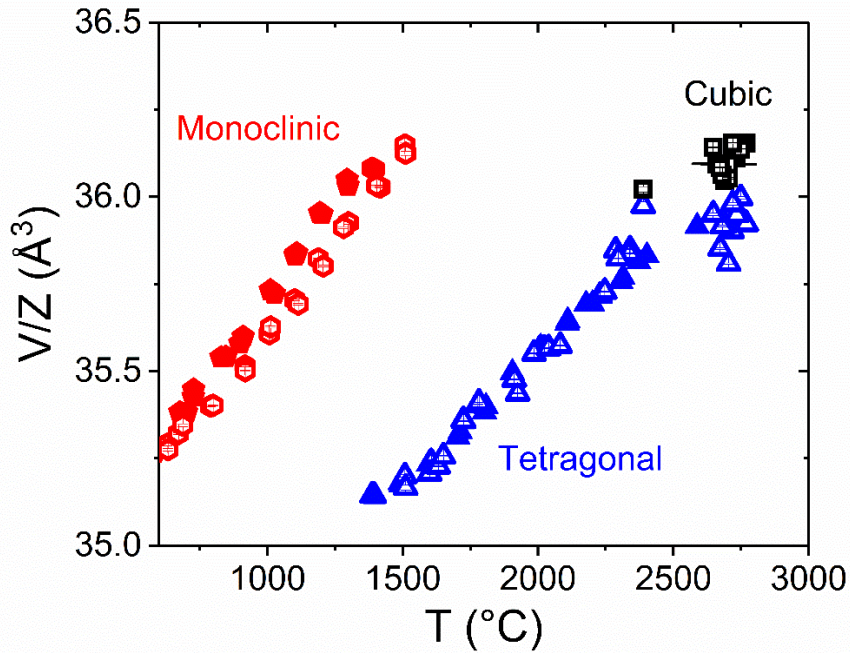

FIG. A1. Unit cell volume of the monoclinic (pentagon), tetragonal (triangles) and cubic (squares) forms of hafnia measured in Argon (closed symbols) and Oxygen (open symbols) atmospheres using high energy X-ray diffraction.

## B. Potential energy surface and descriptor.

For the machine learning of potential energy surface we use the Gaussian Approximation Potential (GAP) and Smooth Overlap of Atomic Position (SOAP) descriptor to construct the similarity kernel described below [14,29]. The total energy of the system is expanded as sum of local energies given by two-body term and many body terms given by nonlinear kernel function.

$$E = \sum_{i<j} U^{(2)}(r_{ij}) + \sum_i \sum_s^M \alpha_s K(d_i, d_s) \quad (B1)$$

The $U^{(2)}$ is a pair-potential and $r_{ij}$ is the distance between atom i and j. The kernel, K provide the similarity measure between two atomic neighborhoods $d_i$ and $d_s$. Chemical descriptor, $d_i$ describes the local chemical environment surrounding the $i^{th}$ atom which is defined by a neighbor density, $\rho_i(r)$ at each point in space r for each atom i. The neighbor density in SOAP kernel is defined as below:

$$\rho_i(r) = \sum_{i'} f_{cut}(r_{ii'}) \, e^{\left[-\frac{(r-r_{ii'})^2}{2\sigma_{at}^2}\right]} \quad (B2)$$

Beyond a cutoff radius $r_{cut}$, $f_{cut}$ is a cutoff function that smoothly goes to zero. $\sigma_{at}$ is a smearing parameter. In practice, the neighbor density is expanded in a local basis of orthogonal radial function, $g_n(r)$ and spherical harmonics, $Y_{lm}(\hat{r})$ given by

$$\rho_i(r) = C_{nlm}^i g_n(r) Y_{lm}(\hat{r}) \quad (B3)$$

With the definition of neighbor density, the SOAP kernel can be estimate as below

$$K(\rho_{SOAP}, \rho'_{SOAP}) = \sum_{n,n',l} P_{n,n',l} P'_{n,n',l} \quad (B4)$$

with the rotationally-invariant power spectrum,

$$P_{n,n',l} = \sum_m C_{nlm}^{i*} C_{nlm}^i \quad (B5)$$

## C. GAP training dataset and parameters.

| Dataset | Temperature (°C) | Number of samples |
|---|---|---|
| Liquid | 3326.85 | 283 |
| Amorphous | 226.85 - 2526.85 | 343 |
| Monoclinic[a] | 25 – 1500 | 428 |
| Tetragonal | 1440 – 2400 | 461 |
| Cubic | 2300 – 2900 | 538 |
| Total | | **2053** |

[a] Includes point defects.

TABLE C1: GAP Training Dataset

| Parameter | SOAP | Two body |
|---|---|---|
| Cut off radius (Å) | 4.0 | 4.0 |
| Smooth cut-off transition (Å) | 1.0 | 1.0 |
| Sparse method | Cur points | uniform |
| Sparse points | 2600 | 40 |
| ($n_{max}$; $l_{max}$) | (6,6) | - |
| Kernel exponent | 4 | - |
| GAP Version | 1548461341 | - |

TABLE C2. GAP Training Parameters

### D. Clustering based active learning scheme.

The pseudocode for the clustering based active learning [15] is reproduced below:

**Initialization:** Trajectory, distance measure, target accuracy.
[1] Input trajectory is converted in to a distance matrix.
[2] Perform unsupervised clustering based on HDBSCAN algorithm to extract uncorrelated clusters.
[3] Training and test configurations are sequentially drawn from the clusters.
[4] Perform hyperparameter search using Bayesian optimization and extract the best model for the chosen training configuration as validated against an independent test configuration.
[5] Draw more samples from the cluster if the required accuracy has not been achieved and repeat [3-5].
[6] Exit if the target accuracy is achieved.

For the accuracy we use Mean Absolute Error (MAE) in GAP predicted Energy in the unit of meV/atom. By default, the workflow uses a target accuracy of 2 meV/atom. For the distance measure, we employ Root Mean Square Deviation of Atomic Positions (RMSD). Given two sets of atomic positions, A and B with 'n' points each. The RMSD[6] is defined as:

$$RMSD(A,B) = \sqrt[2]{\frac{\sum_{i=1}^{n}[(A_{ix}-B_{ix})^2+(A_{iy}-B_{iy})^2+(A_{iz}-B_{iz})^2]}{n}} \qquad (C1)$$

## E. GAP vs Broglia [51] vs COMB [50] inter-atomic potentials

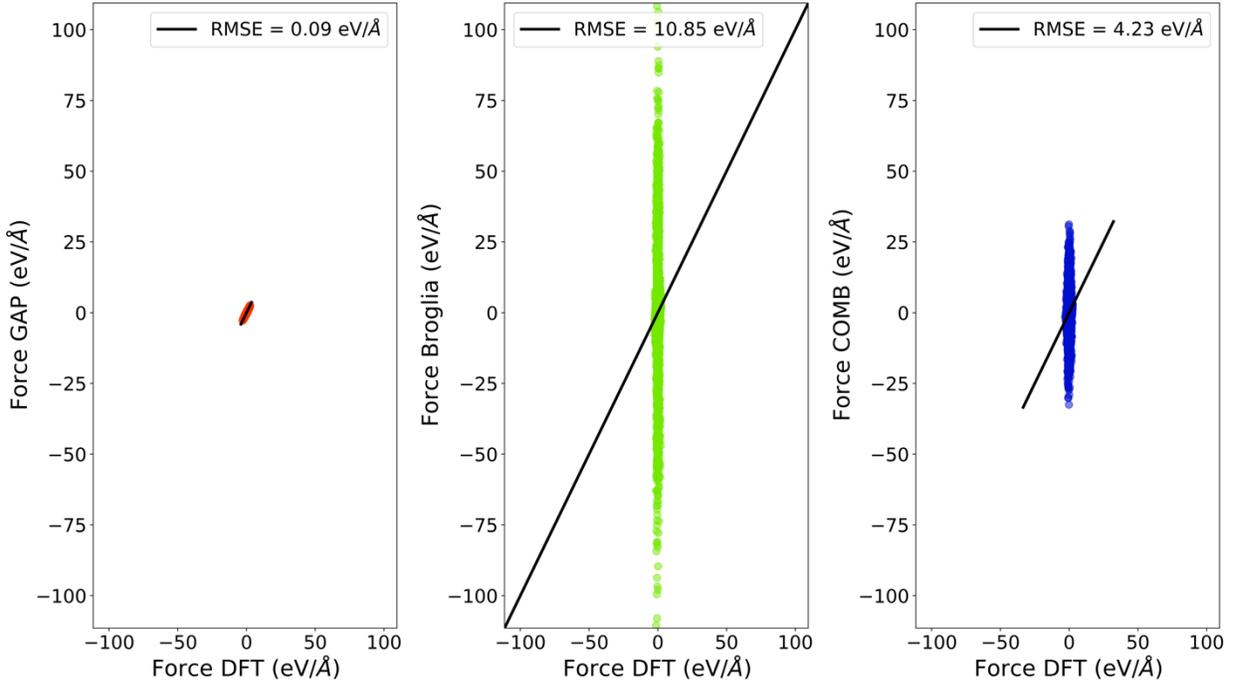

FIG. E1. Comparison of (left panel) DFT force vs GAP force plot, (middle panel) DFT force vs the Broglia et al. [51] force plot and (right panel) DFT force vs COMB [50] potential force plot. 50 randomly chosen configurations from the m-HfO$_2$ AIMD trajectory (i.e. outside of GAP training) is used for this validation test. Root Mean Square Error (RMSE) fit is chosen as force validation metric. The axes scale has been kept consistent for comparison.

## F. Computation of pair distribution functions and structure factor.

The partial pair distribution function (PDF) for atom of species 'j' within a distance of 'r' from atom of species 'i' in the simulation cell can be computed as,

$$g_{ij}(r) = \frac{1}{N_i N_j} \sum_{a=1}^{N_i} \sum_{b=1}^{N_j} \langle \delta(|r_{ab}| - r_{max}) \rangle \quad (F1)$$

The $r_{max}$ is taken as half the simulation box length.

The corresponding partial structure factor is computing through Fourier transform of the partial PDF.

$$S_{ij} = 1 + 4\pi\rho \int_0^\infty r^2 \frac{\sin(Qr)}{Qr}\big(g_{ij}(r) - 1\big) dr \quad (F2)$$

where Q is the scattering vector.

The total structure factor is computed as the weighted sum of partial structure factors,

$$S(Q) = \sum_{j=1}^{N} w_{ij}\, S_{ij} \quad (F3)$$

where $w_{ij}$ is the weight factor is determined from the nature of the incident radiation (i.e. X-ray, neutron etc) and the material composition.

The inverse Fourier transform of the S(Q) gives the total PDF,

$$G(r) - 1 = \frac{1}{2\pi^2 r\rho} \int_0^\infty Q[S(Q) - 1]\sin(Qr)\, dQ \quad (F4)$$

## G. Computed pair distribution functions for pure phases.

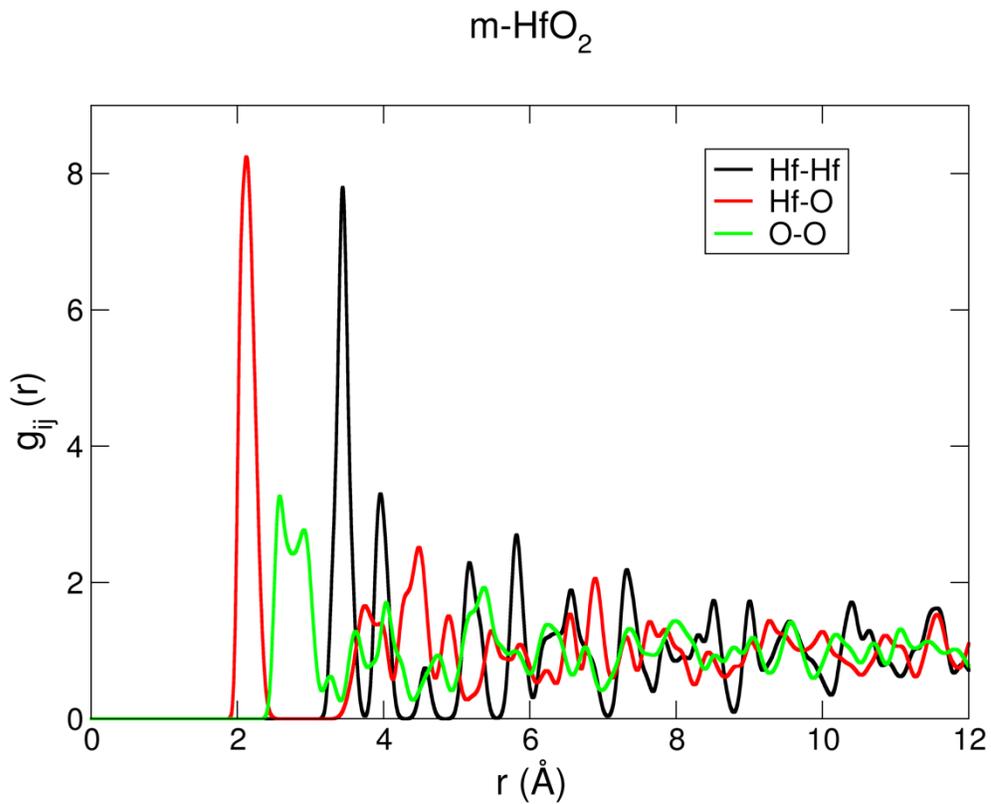

FIG. F1: GAP MD computed PDF for m-$HfO_2$ at 25°C.

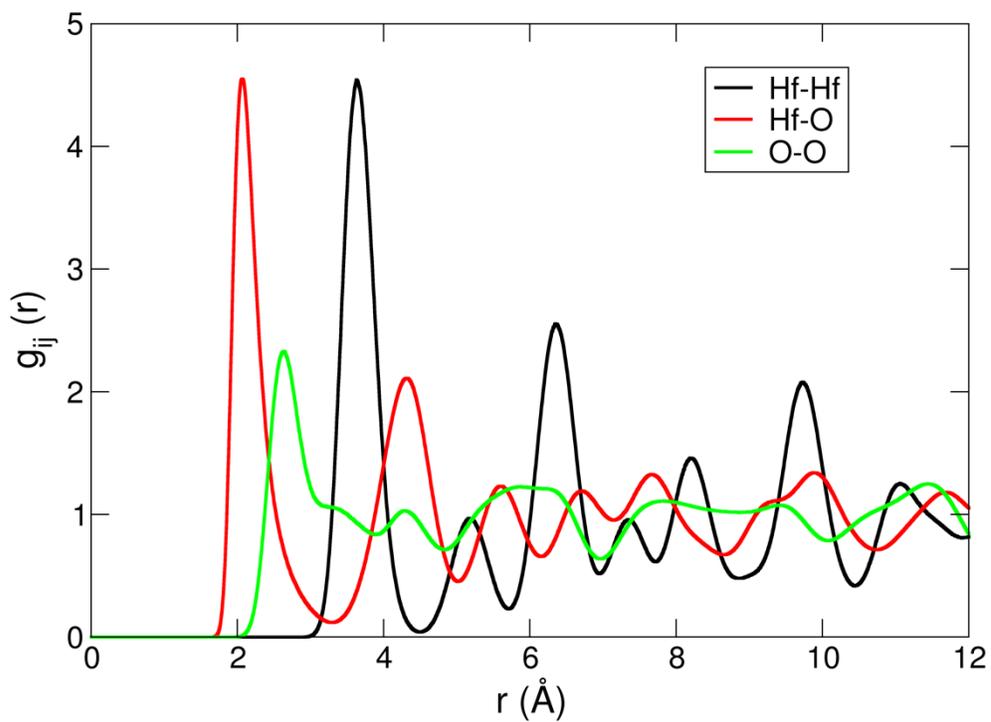

FIG. F2. GAP MD computed PDF for t-HfO$_2$ at 1850°C.

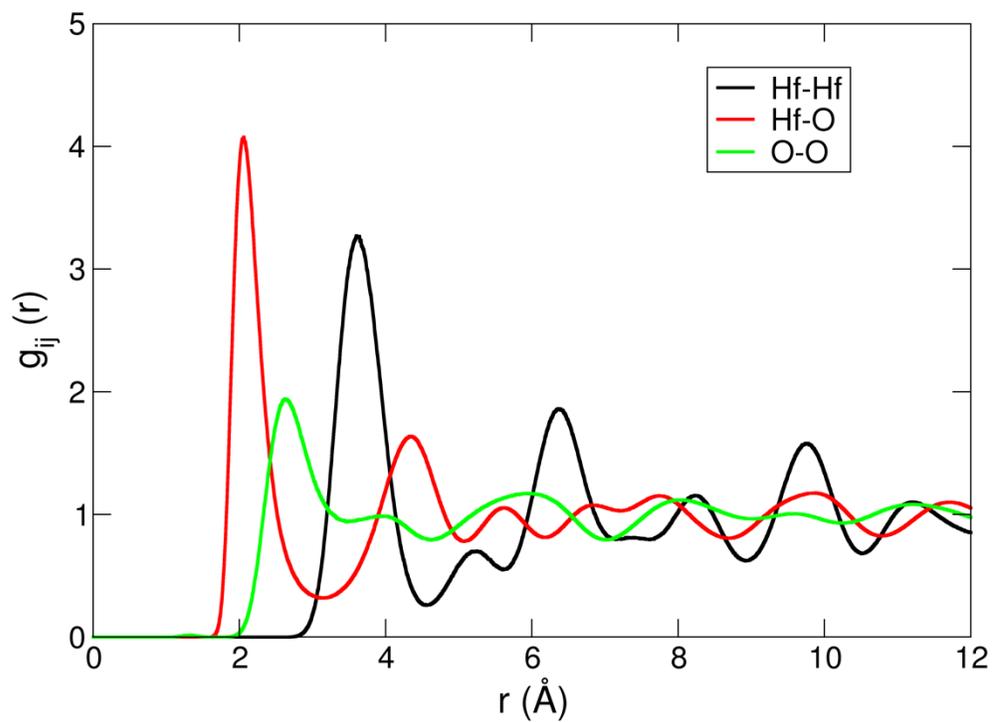

FIG. F3. GAP MD computed PDF for c-HfO$_2$ at 2800°C.

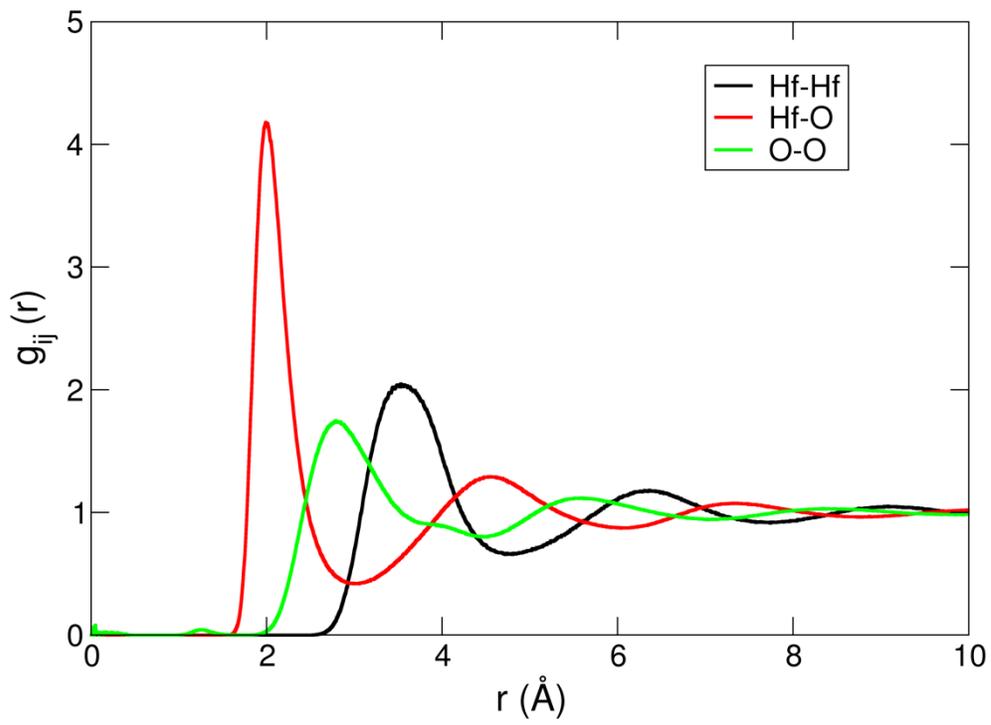

FIG. F4. GAP MD computed PDF for liquid $HfO_2$ at 2900°C.

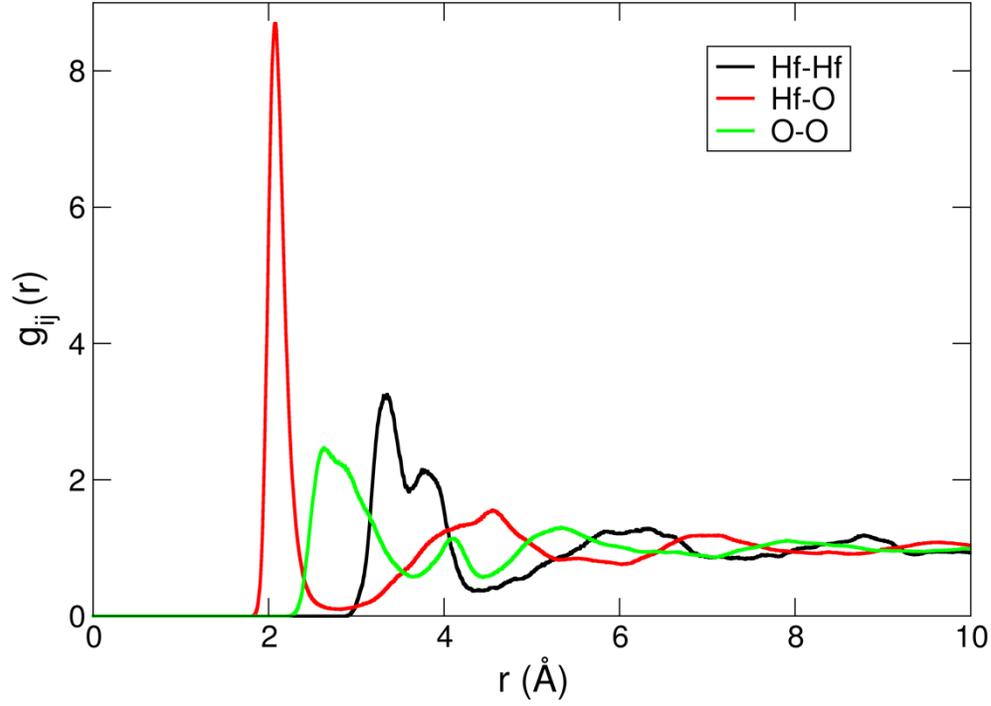

FIG. F5: GAP MD computed PDF for amorphous $HfO_2$.

**H. Computation of X-ray intensity**

The X-ray diffraction intensity at each reciprocal lattice point is computed as a product of structure factor (S) with its complex conjugate ($S^*$) normalized by the number of atoms in the simulation [49].

$$I(Q) = Lp(\theta)\frac{S^*(Q)S(Q)}{N} \qquad (H1)$$

where the Lorentz-polarization factor is given by

$$Lp(\theta) = \frac{1+ cos^2(2\theta)}{cos(\theta)sin^2(\theta)} \qquad (H2)$$

$\theta$ is the scattering angle.

$$S(Q) = \sum_{j=1}^{N} w_j(\theta)\, e^{-2\pi i Q.R_j} \qquad (H3)$$

Where $R_j$ is position of the atom 'j' and $w_j$ the corresponding atomic scattering factor.

With the simulated wavelength $\lambda$,

$$\frac{sin(\theta)}{\lambda} = \frac{|Q|}{2} \qquad (H4)$$